\documentstyle[aaai]{article}

\begin{document}

\title{XNMR -- A tool for knowledge bases exploration}

\author{Lu\'\i s Castro \and David S. Warren\\
Department of Computer Science \\ SUNY at Stony Brook \\
Stony Brook, NY 11794-4400, USA \\ \{luis, warren\}@cs.sunysb.edu} 

\maketitle

\begin{abstract}
\noindent 
XNMR is a system designed to explore the results of combining the
well-founded semantics system XSB with the stable-models evaluator
SMODELS. Its main goal is to work as a tool for fast and interactive
exploration of knowledge bases. 
\end{abstract}

\section{General Information}

The XNMR package is an attempt of integration between the well-founded
semantics system XSB\cite{SSW93a,SaSW96} and the stable models evaluator
SMODELS\cite{NiSi96}. It works in UNIX platforms (eg. Linux, Solaris),
and work is being done to port it to Windows NT.

The package consists of three layers. The bottom layer is a low-level
interface between XSB and SMODELS. This interface is written in C, and
consists of 177 lines of code.

The second layer is a high-level library of Prolog predicates used to
communicate with SMODELS. It has 117 lines of Prolog code.

The third layer is the top-loop evaluator. It is also written in
Prolog, and consists of 421 lines of code.


\section{Description of the System} 

XNMR is a package intended to integrate the possible-worlds stable
semantics of SMODELS with the more skeptical well-founded semantics of
XSB.

It is aimed at allowing the exploration of knowledge bases. It is
well-suited for debugging large knowledge bases, by exploiting the
modularity of such designs.

The package uses XSB as a pre-processing phase, where the well-founded
semantics of a program, with respect to a given query, is computed. As
a result of this process, a residual program is computed, where the
interdependencies of the undefined objects is represented. The
residual program is, then, given as input to SMODELS, so that the
stable models for these undefined objects can be computed.

\section{Applying the System}


\subsection{Methodology}

This work is exploratory, in the sense that it combines
characteristics of two well known systems for logic programming in a
general way. We believe the system may be applied to several
situations where the stable models semantics is preferred, as long as
the notion of relevancy to a query (see \ref{semantics}) is satisfied.

More specifically, one application of this system is in debugging
large knowledge bases. The system can take advantage of its ability to
deal with partial information to allow for debugging of separate
parts, or modules, of a knowledge base. 

Work is in progress to develop a programming methodology that
will allow problems to be described in such a way as to be suitable
for evaluation with XNMR. The methodology is based on ways to encode
the problem such as to avoid the non-relevancy problem.

                                        

\subsection{Specifics}
\label{semantics}

XNMR computes the partial stable models which are total on the
relevant program, given a query. This is due to the combination of a
query-driven system (XSB) to a bottom-up evaluator like SMODELS. 

The models computed are total stable models with respect to the
relevant parts of the program, according to the given query.

We believe this system may have successful applications on areas like
debugging of knowledge bases, and null-valued databases.


\subsection{Users and Useability}

The system is to be used by people with an understanding of the issues
involved in the combination of well-founded and stable models
semantics. Since it is integrated with XSB Prolog, it is general
enough to be used in several applications.

Even though the system is not, at the moment, being used outside our
research group, there are already people interested in applying it to
several applications.


\section{Evaluating the System}


\subsection{Benchmarks}

To our knowledge, there are no comparative systems available
today. One reasonable way to benchmark the systems, though, would be
considering XSB as a pre-processor to SMODELS. This way, accepted
benchmarks for SMODELS could be applied to the system, as long as they
were well-behaved with respect to the relevant-program restriction
imposed by the top-down nature of XSB evaluation.

The system is incorporated to the XSB Prolog system, therefore having
the same user-friendliness associated to such systems.


\subsection{Comparison} 

The system is inherently logic-based. As stated in the previous
section, there are no standard benchmarks to the system, due to its
unique combination of top-down and bottom-up evaluators, which results
in a restricted semantics being computed. We believe such benchmarks
should be considered in an applications-based framework.


\subsection{Problem Size} 

The XNMR system, by itself, doesn't impose any restrictions to the
size of problems being handled. So, the restrictions of the system are
those of XSB and SMODELS. We also note that the system is not a
prototype, and has been fully implemented as a package of XSB.


\bibliographystyle{aaai}
\bibliography{manual}

\end{document}